\documentclass[prb,twocolumn,superscriptaddress,showpacs,preprintnumbers,amsmath,amssymb]{revtex4-1}

\usepackage{graphicx}% Include figure files
\usepackage{dcolumn}% Align table columns on decimal point
\usepackage{bm}% bold math
\usepackage{color,ulem}% color and ulem

\begin{document}

%\preprint{APS/PRB}

\title{Coexisting spin resonance and long-range magnetic order of Eu in EuRbFe$_4$As$_4$}% Force line breaks with \\

\author{K.~Iida}\email{k\_iida@cross.or.jp}
\affiliation{Neutron Science and Technology Center, Comprehensive Research Organization for Science and Society (CROSS), Tokai, Ibaraki 319-1106, Japan}

\author{Y.~Nagai}
\affiliation{CCSE, Japan Atomic Energy Agency (JAEA), Kashiwa, Chiba 277-0871, Japan}
\affiliation{Mathematical Science Team, RIKEN Center for Advanced Intelligence Project (AIP), 1-4-1 Nihonbashi, Chuo-ku, Tokyo 103-0027, Japan}

\author{S.~Ishida}
\affiliation{National Institute of Advanced Industrial Science and Technology (AIST), Tsukuba, Ibaraki 305-8568, Japan}

\author{M.~Ishikado}
\affiliation{Neutron Science and Technology Center, Comprehensive Research Organization for Science and Society (CROSS), Tokai, Ibaraki 319-1106, Japan}

\author{N.~Murai}
\affiliation{Materials and Life Science Division, J-PARC Center, Tokai, Ibaraki 319-1195, Japan}

\author{A.~D.~Christianson}
\affiliation{Neutron Scattering Division, Oak Ridge National Laboratory, Oak Ridge, Tennessee 37831, USA}
\affiliation{Materials Science and Technology Division, Oak Ridge National Laboratory, Oak Ridge, Tennessee 37831, USA}

\author{H.~Yoshida}
\affiliation{Department of Physics, Faculty of Science, Hokkaido University, Sapporo, Hokkaido 060-0810, Japan}

\author{Y.~Inamura}
\affiliation{Materials and Life Science Division, J-PARC Center, Tokai, Ibaraki 319-1195, Japan}

\author{H.~Nakamura}
\affiliation{CCSE, Japan Atomic Energy Agency (JAEA), Kashiwa, Chiba 277-0871, Japan}

%\author{T.~Matsukawa}
%\affiliation{Frontier Research Center for Applied Atomic Sciences, Ibaraki University, Tokai, Ibaraki 319-1106, Japan}

%\author{A.~Hoshikawa}
%\affiliation{Frontier Research Center for Applied Atomic Sciences, Ibaraki University, Tokai, Ibaraki 319-1106, Japan}

%\author{T.~Ishigaki}
%\affiliation{Frontier Research Center for Applied Atomic Sciences, Ibaraki University, Tokai, Ibaraki 319-1106, Japan}

%\author{Y.~Kawamura}
%\affiliation{Neutron Science and Technology Center, Comprehensive Research Organization for Science and Society (CROSS), Tokai, Ibaraki 319-1106, Japan}

\author{A.~Nakao}
\affiliation{Neutron Science and Technology Center, Comprehensive Research Organization for Science and Society (CROSS), Tokai, Ibaraki 319-1106, Japan}

\author{K.~Munakata}
\affiliation{Neutron Science and Technology Center, Comprehensive Research Organization for Science and Society (CROSS), Tokai, Ibaraki 319-1106, Japan}

\author{D.~Kagerbauer}
\affiliation{Atominstitut, TU Wien, Stadionalle 2, 1020 Vienna, Austria}

\author{M.~Eisterer}
\affiliation{Atominstitut, TU Wien, Stadionalle 2, 1020 Vienna, Austria}

\author{K.~Kawashima}
\affiliation{National Institute of Advanced Industrial Science and Technology (AIST), Tsukuba, Ibaraki 305-8568, Japan}
\affiliation{IMRA Material R\&D Co., Ltd., Kariya, Aichi 448-0032, Japan}

\author{Y.~Yoshida}
\affiliation{National Institute of Advanced Industrial Science and Technology (AIST), Tsukuba, Ibaraki 305-8568, Japan}

\author{H.~Eisaki}
\affiliation{National Institute of Advanced Industrial Science and Technology (AIST), Tsukuba, Ibaraki 305-8568, Japan}

\author{A.~Iyo}
\affiliation{National Institute of Advanced Industrial Science and Technology (AIST), Tsukuba, Ibaraki 305-8568, Japan}

\date{\today}% It is always \today, today,

\begin{abstract}
Magnetic excitations and magnetic structure of EuRbFe$_4$As$_4$ were investigated by inelastic neutron scattering (INS), neutron diffraction, and random phase approximation (RPA) calculations.
Below the superconducting transition temperature $T_\text{c}=36.5$~K, the INS spectra exhibit the neutron spin resonances at $Q_\text{res}=1.27(2)$~$\text{\AA}^{-1}$ and $1.79(3)$~$\text{\AA}^{-1}$.
They correspond to the $\mathbf{Q}=(0.5,0.5,1)$ and $(0.5,0.5,3)$ nesting wave vectors, showing three dimensional nature of the band structure.
The characteristic energy of the neutron spin resonance is $E_\text{res}=17.7(3)$~meV corresponding to $5.7(1)k_\text{B}T_\text{c}$.
Observation of the neutron spin resonance mode and our RPA calculations in conjunction with the recent optical conductivity measurements are indicative of the $s_\pm$ superconducting pairing symmetry in EuRbFe$_4$As$_4$.
In addition to the neutron spin resonance mode, upon decreasing temperature below the magnetic transition temperature $T_\text{N}=15$~K, the spin wave excitation originating in the long-range magnetic order of the Eu sublattice was observed in the low-energy inelastic channel.
Single-crystal neutron diffraction measurements demonstrate that the magnetic propagation vector of the Eu sublattice is $\mathbf{k}=(0, 0, 0.25)$, representing the three-dimensional antiferromagnetic order.
Linear spin wave calculations assuming the obtained magnetic structure with the intra- and inter-plane nearest neighbor exchange couplings of $J_1/k_\text{B}=-1.31$~K and $J_c/k_\text{B}=0.08$~K can reproduce quantitatively the observed spin wave excitation.
Our results show that superconductivity and long-range magnetic order of Eu coexist in EuRbFe$_4$As$_4$ whereas the coupling between them is rather weak.
\end{abstract}

%\pacs{74.70.Pq, 78.70.Nx}% PACS, the Physics and Astronomy
\maketitle

\section{Introduction}
Recently, new family of iron-based superconductors $AeA$Fe$_4$As$_4$ with $Ae=$ Ca, Sr, Eu and $A=$ K, Rb, Cs (termed ``1144'') were reported by Iyo \textit{et} \textit{al}~\cite{CaKFe4As4_powder}.
EuRbFe$_4$As$_4$~\cite{EuRbFe4As4_1,EuRbFe4As4_2} can be regarded as an intergrowth of EuFe$_2$As$_2$ and RbFe$_2$As$_2$, both of which belong to so-called ``122'' family.
Unlike solid solutions such as (Eu$_{1-x}$Na$_x$)Fe$_2$As$_2$~\cite{EuNaFe2As2_1,EuNaFe2As2_2} and (Eu$_{1-x}$La$_x$)Fe$_2$As$_2$~\cite{EuLaFe2As2_1}, in EuRbFe$_4$As$_4$, Eu$^{2+}$ and Rb$^+$ ions occupy crystallographically inequivalent sites alternately between the FeAs layers along the $c$ axis because of the large difference between their ionic radii~\cite{EuRbFe4As4_4}.
Consequently, EuRbFe$_4$As$_4$ crystallizes in the tetragonal symmetry ($P4/mmm$), and no structural phase transition takes place down to 2~K~\cite{EuRbFe4As4_4}.
The unique structure of the 1144 compounds is that there are two different kinds of As sites in contrast to the single As site in the 122 compounds.

As is well established, the non-superconducting parent compound EuFe$_2$As$_2$ undergoes a spin-density-wave (SDW) order in the Fe sublattice accompanied by a tetragonal-to-orthorhombic structural phase transition below 190~K~\cite{EuFe2As2_1,EuFe2As2_2}.
Hole doping by substitution of Rb$^+$ for Eu$^{2+}$ in EuFe$_2$As$_2$ gives rise to suppression of the SDW order in EuRbFe$_4$As$_4$, in favor of the occurrence of superconductivity below $T_\text{c}=36.5$~K~\cite{EuRbFe4As4_1,EuRbFe4As4_2}.
Bulk property measurements indicated that EuRbFe$_4$As$_4$ shows behaviors similar to optimally doped 122 superconductors~\cite{EuRbFe4As4_1,EuRbFe4As4_7}.
The normal state of the FeAs layer in EuRbFe$_4$As$_4$ is paramagnetic metal~\cite{EuRbFe4As4_5}.
On the other hand, Curie-Weiss fitting on the magnetic susceptibility measurement suggests that Eu$^{2+}$ ion has a local magnetic moment with $S=7/2$ and the Curie-Weiss temperature $\theta$ is positive, suggesting that the dominant magnetic interaction between Eu$^{2+}$ ions is ferromagnetic~\cite{EuRbFe4As4_2,EuRbFe4As4_4,EuRbFe4As4_7,EuRbFe4As4_3}.
Upon further cooling below $T_\text{c}$, the Eu sublattice shows a long-range magnetic order at $T_\text{N}=15$~K~\cite{EuRbFe4As4_2,EuRbFe4As4_7}.
The Eu sublattice of parent EuFe$_2$As$_2$ undergoes a so-called $A$-type antiferromagnetic order below $T_\text{N}=19$~K~\cite{EuFe2As2_6,EuFe2As2_7}, and smaller $T_\text{N}$ in EuRbFe$_4$As$_4$ is due to the longer interlayer distance between the Eu layers compared to EuFe$_2$As$_2$.
In fact, upon increasing hydrostatic pressure, the lattice constant $c$ (or interlayer distance of the Eu sublattice) decreases while $T_\text{N}$ increases in EuRbFe$_4$As$_4$~\cite{EuRbFe4As4_6,EuRbFe4As4_11}.
It is noticeable that no reentrant superconductivity was observed in EuRbFe$_4$As$_4$~\cite{EuRbFe4As4_1,EuRbFe4As4_2} unlike observed in Eu(Fe$_{0.89}$Co$_{0.11}$)$_2$As$_2$~\cite{EuFeCo2As2_1}, Eu(Fe$_{0.86}$Ir$_{0.14}$)$_2$As$_2$~\cite{EuFeIr2As2_1}, Eu(Fe$_{0.75}$Ru$_{0.25}$)$_2$As$_2$~\cite{EuFeRu2As2_1}, Eu(Fe$_{0.93}$Rh$_{0.07}$)$_2$As$_2$~\cite{EuFeRh2As2_1}, and EuFe$_2$(As$_{0.7}$P$_{0.3}$)$_2$~\cite{EuFe2AsP2_1}.

Although $s$-wave superconductivity with a complete gap throughout the Fermi surface is reported by the optical conductivity measurements~\cite{EuRbFe4As4_8}, the microscopic mechanism underlying the superconductivity in EuRbFe$_4$As$_4$ has not been completely elucidated yet.
In particular, due to the large neutron absorption cross section of Eu, the neutron spin resonance~\cite{Review_2,Review_3,Review_4,Resonance1,Resonance2,Resonance3,Resonance4,Resonance5,Resonance6,Resonance7}, which provides a strong constraint on the relative signs of the superconducting gaps on different portions of the Fermi surfaces~\cite{spm_1,spm_2,cuprate_1,cuprate_2,cuprate_3,cuprate_4,cuprate_5,HeavyFermion_1}, has not been reported.
In the meanwhile, the detailed magnetic structure of the Eu sublattice in EuRbFe$_4$As$_4$ is not determined yet using a microscopic technique such as neutron diffraction.
Investigation of the magnetic structure by means of the microscopic technique is necessary because various types of the magnetic structures of the Eu sublattice are reported in the doped EuFe$_2$As$_2$ compounds~\cite{EuFeCo2As2_5,EuFeIr2As2_3,EuFe2AsP2_11,EuFe2As2_6,EuFe2As2_7}.
Moreover, there is no experimental report on the spin wave excitation from the long-range order of Eu$^{2+}$ in EuRbFe$_4$As$_4$.

So far, there is no experimental evidence about the sign-changing superconductivity in Eu-containing iron-based superconductors, which can demonstrate that Eu-containing iron-based superconductors also belong to the same paradigm as other iron-based superconductors in spite of the local magnetic moments of Eu$^{2+}$ between FeAs layers.
Moreover, there is a longstanding issue about the possible interplay between the long-range magnetic order of Eu$^{2+}$ local moments and the superconductivity of the FeAs layer in the Eu-containing iron-based superconductors~\cite{EuFeIr2As2_1,EuFeIr2As2_3,coupling_1,coupling_2,coupling_3,coupling_4,coupling_5,coupling_6,coupling_7,coupling_8,coupling_9,coupling_10}.
It is, therefore, important to report such coupling in the case of EuRbFe$_4$As$_4$ since EuRbFe$_4$As$_4$ contains no substitutional disorder.
Measurements of magnetic excitations from the FeAs and Eu layers in  $(Q, \hbar\omega)$ space in EuRbFe$_4$As$_4$ will give us the key ingredient to discuss the possible cooperative behaviors.
In this paper, we investigated the neutron spin resonance by combination of inelastic neutron scattering (INS) measurements and random phase approximation (RPA) calculations.
In addition, we also measured spin wave excitation and single-crystal neutron diffraction patterns to determine the spin Hamiltonian and the magnetic structure of the Eu sublattice.
Finally, we will discuss the effect of the magnetic long-range order on the superconductivity in EuRbFe$_4$As$_4$.

\section{Experimental details and theoretical methods}
The details of the synthesis of EuRbFe$_4$As$_4$ is described in elesewhere~\cite{CaKFe4As4_powder,EuRbFe4As4_1}, and a magnetic susceptibility measurement exhibits $T_\text{c}=36.5$~K and $T_\text{N}=15$~K in the current polycrystalline sample as shown in Supplemental Material~\cite{Supplement}.
Polycrystalline EuRbFe$_4$As$_4$ with a mass of $\sim4$~g was used for our INS measurements.
We performed high- and low-energy INS measurements using two Fermi-chopper spectrometers ARCS at Spallation Neutron Source (SNS)~\cite{ARCS} and 4SEASONS at Japan Proton Accelerator Research Complex (J-PARC)~\cite{4SEASONS_1,4SEASONS_2}.
Neutron incident energies $E_\text{i}=70.0$~meV were used at ARCS while $E_\text{i}=8.5$~meV at 4SEASONS, and energy resolutions (full width at half maximum) at the elastic channel were 3.2 and 0.36~meV, respectively.
INS data were analyzed by the software suites Utsusemi~\cite{4SEASONS_3}.
To reduce the effect of neutron absorption by Eu, thin-plate cans whose effective thickness were 1~mm (ARCS) and 0.5~mm (4SEASONS) were used.
For the absolute intensity scale in the present INS measurements, incoherent scattering from EuRbFe$_4$As$_4$ was used after correction for neutron absorption~\cite{AbsoluteNormalization}.
Single-crystal neutron diffraction measurements on EuRbFe$_4$As$_4$ were also carried out using the time-of-flight (TOF)  single-crystal neutron diffractometer SENJU~\cite{SENJU} in J-PARC.
A single crystal with dimension about $1\times1\times0.1$~mm$^3$ was grown for neutron diffraction measurements, and the obtained lattice constants at 20.1~K are $a=3.85923(2)$ and $c=13.18563(7)~\text{\AA}$ for the space group $P4/mmm$.
%3.94~g for 4SEASONS
%0.83~g for iMATERIA

The first-principles calculation package employed is VASP\cite{Kresse1,Kresse2}, which adopts the PAW method\cite{Blochl1,Blochl2} and the GGA exchange-correlation energy\cite{Perdew}. 
The number of ${\bf k}$ points is taken as $16 \times 16 \times 5$, and self-consistent loops are repeated until the energy deviation becomes less than $10^{-6}$~eV, with the cutoff energy being 500~eV. 
The density functional theory often fails to handle $f$ orbitals correctly due to self-interaction error of localized $f$-electrons. To avoid this failure, localized 4$f$ electrons in Eu atoms are treated as core electrons in our calculations.
In the RPA calculations, for the effective model, maximally localized Wannier functions from twenty Fe $3d$ bands are constructed\cite{Mostofi}. 
We employ the orbital-interaction coefficient $U_s = a U^{rs}_{qt,M}$ obtained by first-principles calculation for the 122 system\cite{Miyake} and we put $a = 1.7$.
The number of ${\bf k}$ point is $96 \times 96 \times 8$ and the smearing factor is $\eta = \Delta_0/8$ where $\Delta_0$ denote superconducting gap.
We take $\Delta_{0}$ ($-\Delta_0$) on the hole (electron) Fermi surfaces around $\Gamma$ ($M$) point. 
We set $\Delta_0=0.1$~eV to avoid numerical difficulty. 
We introduce a Gaussian cutoff for the gap $\Delta^{\nu}_{k} = \Delta^{\nu} \exp[-\epsilon_k^{\nu}/(4\Delta_0)]$ with the normal-state energy dispersion $\epsilon_k^{\nu}$ on the $\nu$-th band \cite{NagaiKuroki}.

Linear spin wave (LSW) calculations were performed using the SpinW software~\cite{SpinW}.
The squared magnetic form factor of Eu$^{2+}$ ion~\cite{InternationalTalbes} and $\hbar\omega$-dependent energy resolution for $E_\text{i}=8.5$~meV at 4SEASONS~\cite{4SEASONS_4} were included in the LSW calculations.

\section{Results and Discussion}
\begin{figure}[t]
\centering
\includegraphics[width=6.97cm]{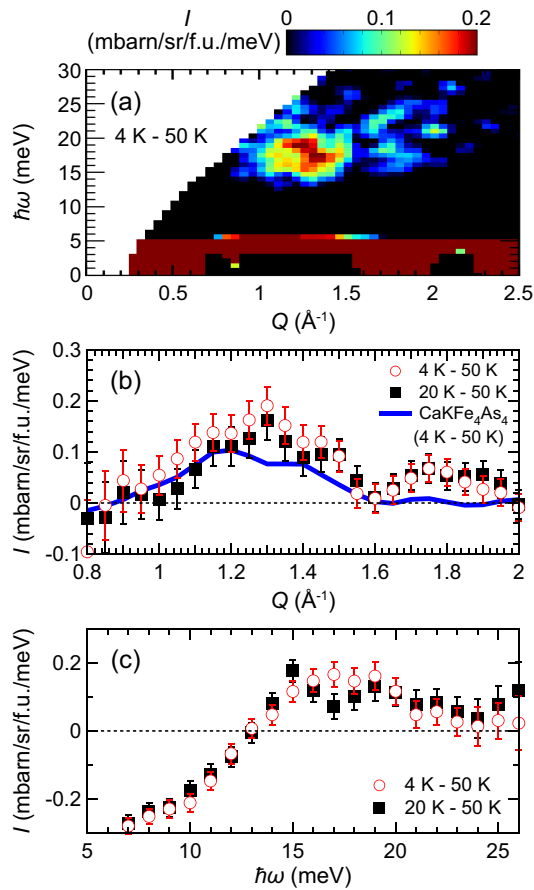}
\caption{
High-energy INS results on EuRbFe$_4$As$_4$ measured at ARCS using $E_\text{i}=70.0$~meV.
In all panels, the temperature differences between $T=4$ (or 20) and 50~K are shown.
(a) Neutron scattering intensity map $I(Q,\hbar\omega)$.
(b) $Q$ dependences of the neutron scattering intensities with the energy window of $\hbar\omega=[15,20]$~meV.
Solid line represents the $Q$ dependence of the neutron spin resonance mode with $\hbar\omega=[11,16]$~meV in CaKFe$_4$As$_4$ measured at ARCS using $E_\text{i}=70.0$~meV for reference~\cite{CaKFe4As4_Neutron1}.
(c) Energy cuts with $Q=[1.1,1.4]$~$\text{\AA}^{-1}$.
}
\label{Fig:HighE}
\end{figure}

Figure~\ref{Fig:HighE}(a) depicts the neutron scattering intensity ($I$) map from EuRbFe$_4$As$_4$ as a function of momentum ($Q$) and energy $(\hbar\omega)$ transfers.
The data at 50~K was subtracted from that at 4~K.
There is a well localized excitation in both $Q$ and $\hbar\omega$: $1.1\le Q\le1.3~\text{\AA}^{-1}$ and $13\le\hbar\omega\le20~\text{meV}$ which are comparable to the neutron spin resonance in isostructural CaKFe$_4$As$_4$ ($1.0\le Q\le1.4~\text{\AA}^{-1}$ and $9\le\hbar\omega\le16~\text{meV}$)~\cite{CaKFe4As4_Neutron1}.
Since the localized excitation increases in intensity at lower temperature, the observed excitation is magnetic in origin.
To further classify the localized magnetic excitation in EuRbFe$_4$As$_4$, $Q$ and $\hbar\omega$ dependences of the neutron scattering intensities [$I(Q)$ and $I(\hbar\omega)$] are investigated.

\begin{figure}[t]
\centering
\includegraphics[width=5.92cm]{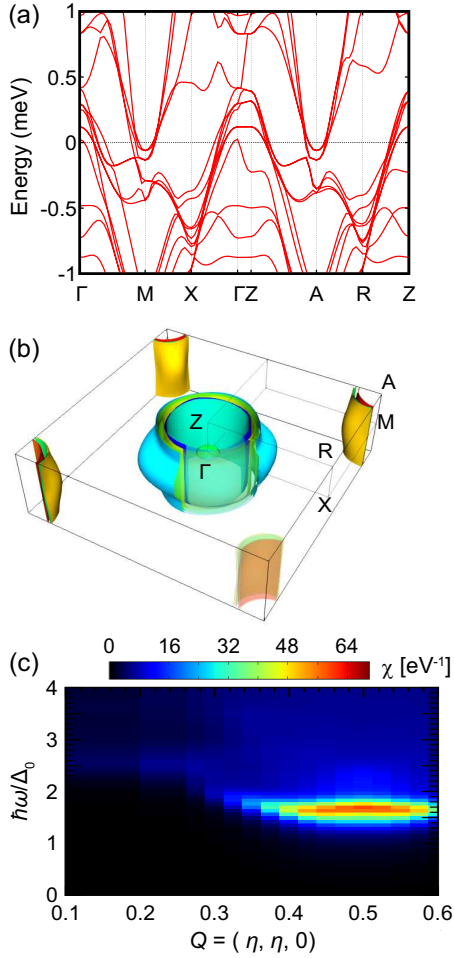}
\caption{
(a) Electronic structures and (b) Fermi surfaces for EuRbFe$_4$As$_4$.
(c) Dynamical spin susceptibility $\chi''(\mathbf{Q},\hbar\omega)$ of EuRbFe$_4$As$_4$ obtained by using the multiorbital RPA with the use of the twenty-orbital three-dimensional tight-binding model.
}
\label{Fig:calculation}
\end{figure}

The temperature difference of $I(Q)$ between 4 and 50~K is shown in Fig.~\ref{Fig:HighE}(b).
$I(Q)$ is averaged over the energy-transfer range of $15\le\hbar\omega\le20$~meV.
A Gaussian function was fitted to the $Q$ cut at 4~K, yielding the peak position of $Q=1.27(2)$~$\text{\AA}^{-1}$.
As in CaKFe$_4$As$_4$~\cite{CaKFe4As4_Neutron1,CaKFe4As4_Neutron2}, the peak position in $Q$ is close to both the $\mathbf{Q}=(0.5,0.5)$ antiferromagnetic wave vector ($1.14~\text{\AA}^{-1}$) and the in-plane propagation vector in the parent compound EuFe$_2$As$_2$ ($1.13$~$\text{\AA}^{-1}$)~\cite{EuFe2As2_2}.
As discussed later, the $(0.5,0.5)$ antiferromagnetic vector is in good condition for the nesting between the electron and hole pockets at the Fermi surface [Fig.~\ref{Fig:calculation}(b)] via spin fluctuations~\cite{spm_1,spm_2} as other iron pnictide systems.
Thus, the observed $Q$ dependence indicates that the dominant magnetic fluctuation of the FeAs layers in EuRbFe$_4$As$_4$ originates in the $(0.5,0.5)$ nesting.
There is nonnegligible signal at higher $Q$, and the Gaussian fitting yields the peak center $Q=1.79(3)$~$\text{\AA}^{-1}$.
Recent single-crystal INS measurements on CaKFe$_4$As$_4$ reported the strong $L$ dependences of the neutron spin resonance modes, representing three dimensional nature of the band structure~\cite{CaKFe4As4_Neutron2}.
The low-energy spin resonance modes in CaKFe$_4$As$_4$ shows the maximum intensity at $\mathbf{Q}=(0.5,0.5,L)$ with odd $L$.
In the case of EuRbFe$_4$As$_4$, $(0.5,0.5,1)$ and $(0.5,0.5,3)$ correspond to 1.25 and 1.84~$\text{\AA}^{-1}$, and the $L$ dependence of the neutron spin resonance, therefore, naturally explains both the first and second peaks observed in the $Q$ cut of EuRbFe$_4$As$_4$.

The temperature difference of $I(\hbar\omega)$ between 4 and 50~K is plotted in Fig.~\ref{Fig:HighE}(c).
$I(\hbar\omega)$ is averaged over $1.1\le Q\le1.4$~$\text{\AA}^{-1}$.
The intensity at 4~K is enhanced over the energy range of $13\le\hbar\omega\le23$~meV, and the spectral-weight gain is transferred by the depletion at lower energies ($\hbar\omega\le13$~meV).
These are the typical behaviors of the neutron spin resonance observed in many iron-based superconductors~\cite{CaKFe4As4_Neutron1,1111_2,Ba122_4,Ba122_1,CaKFe4As4_Neutron2}.
In addition, the scattering intensity of the observed localized mode in EuRbFe$_4$As$_4$ is almost the same as that of the neutron spin resonance in CaKFe$_4$As$_4$~\cite{CaKFe4As4_Neutron1} [see the solid line in Fig.~\ref{Fig:HighE}(b)].
Therefore, the observed localized magnetic excitation can be assigned to be the neutron spin resonance mode.
By fitting a Gaussian function to $I(\hbar\omega)$ at 4~K using the data between 13 and 24~meV, the characteristic energy of the neutron spin resonance is estimated to be $E_\text{res}=17.7(3)$~meV which is equivalent to $5.7(1)k_\text{B}T_\text{c}$, in agreement with the quantitative relationship between $E_\text{res}$ and $T_\text{c}$ widely reported in iron-based superconductors~\cite{Resonance1,Resonance2,Resonance3,CaKFe4As4_Neutron1,1111_2,111_1,Ba122_4,1111_3,Ba122_1,CaKFe4As4_Neutron2,LiODFeSe_1}.
The neutron spin resonance in EuRbFe$_4$As$_4$ is also studied by RPA calculations as in the following text.

We calculate the band structure and the Fermi surfaces of EuRbFe$_4$As$_4$ as shown in Figs.~\ref{Fig:calculation}(a) and \ref{Fig:calculation}(b).
There are six two-dimensional cylindrical and one three-dimensional small hole Fermi surfaces around $\Gamma$ point and four cylindrical electron Fermi surfaces around $M$ point, which are similar to those of CaKFe$_4$As$_4$~\cite{CaKFe4As4_Neutron1}.
We calculate the dynamical spin susceptibility $\chi''(\mathbf{Q},\hbar\omega)$ of EuRbFe$_4$As$_4$ on the basis of multiorbital RPA\cite{NagaiKuroki} with the use of the effective twenty-orbital three-dimensional tight-binding model at $T=0$.
Figure~\ref{Fig:calculation}(c) shows $\eta-\hbar\omega$ dependence of the dynamical spin susceptibility of EuRbFe$_4$As$_4$.
Here, $\eta$ is the parameter which indicates $\mathbf{Q}=(\eta,\eta,0)$.
The resonance peak due to the $s_{\pm}$-wave pairing states occurs around $\eta=0.5$ or $\mathbf{Q}=(0.5,0.5,0)$, which is similar to those in other iron-based superconductors.
The broadness of the peaks in the region $0.4 < \eta < 0.6$ originates from the fact that there are many Fermi surfaces with different sizes.
As mentioned above, the optical conductivity measurements~\cite{EuRbFe4As4_8} reported the $s$-wave superconductivity in EuRbFe$_4$As$_4$.
INS measurements and RPA calculations indicate the sign-changing superconducting gaps on different portions of the Fermi surfaces connected by the $(0.5, 0.5)$ nesting vector~\cite{INStheory_1,INStheory_2}.
Combined with these results, we thus conclude that $s_\pm$ superconducting pairing symmetry is realized in EuRbFe$_4$As$_4$ as in other iron based superconductors.

\begin{figure}[t]
\centering
\includegraphics[width=8.54cm]{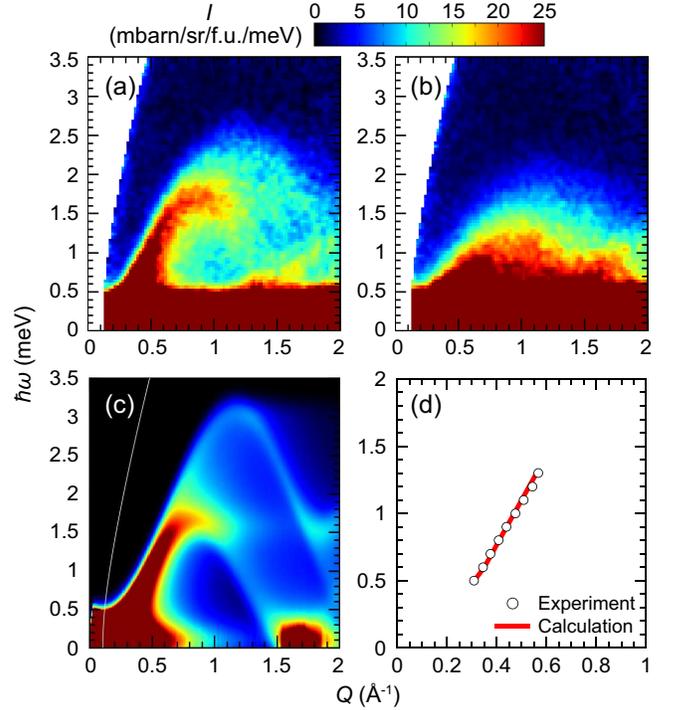}
\caption{
Low-energy INS results of EuRbFe$_4$As$_4$ measured at 4SEASONS with $E_\text{i}=8.5$~meV.
Neutron scattering intensity maps at (a) $T=10.5$ and (b) 25.0~K.
(c) Calculated LSW intensity map of EuRbFe$_4$As$_4$ at 10.5~K.
The magnetic structure depicted in Fig.~\ref{Fig:Diffraction}(c) with the intra- and inter-plane nearest neighbor exchange couplings of $J_1/k_\text{B}=-1.31$~K and $J_c/k_\text{B}=0.08$~K~\cite{EuRbFe4As4_7} was considered.
White line shows the detector coverage boundary of 4SEASONS for $E_\text{i}=8.5$~meV.
(d) Experimental and calculated powder-averaged dispersion relations of the spin wave excitation at 10.5~K in EuRbFe$_4$As$_4$.
Experimental dispersion relation was obtained by fitting $I(Q)$ at each energy window.
}
\label{Fig:LowE}
\end{figure}

In addition to the spin resonance excitation, EuRbFe$_4$As$_4$ shows strong dispersive signals in the low-energy inelastic channel ($\hbar\omega\le5$~meV) at 4~K as we can see in Fig.~\ref{Fig:HighE}(a).
To investigate in detail the excitation, low-energy $I(Q, \hbar\omega)$ maps at $T=10.5$ and 25.0~K are presented respectively in Figs.~\ref{Fig:LowE}(a) and \ref{Fig:LowE}(b).
At 25.0~K ($>T_\text{N}$), a quasielastic excitation was observed [Fig.~\ref{Fig:LowE}(b)].
This is a common feature of the paramagnetic state of correlated localized spins~\cite{Ferromagnetic1}.
Indeed, the signature of the short-range correlation above $T_\text{N}$ is also reported by the heat capacity measurement~\cite{EuRbFe4As4_4,EuRbFe4As4_7}.
In contrast, spin wave dispersion develops at 10.5~K ($<T_\text{N}$) due to the long-range magnetic order of the Eu sublattice [Fig.~\ref{Fig:LowE}(a)].
The magnetic excitation arises from $Q=0$ (or close to $Q=0$), indicating that the dominant magnetic interaction and the corresponding magnetic structure are ferromagnetic as in conventional ferromagnetic spin systems~\cite{Ferromagnetic1}.
To determine the peak position of the spin dispersion relation at low $Q$ and $\hbar\omega$, $I(Q)$ cuts at several energy windows are plotted in Supplemental Material~\cite{Supplement}.
The Warren's function~\cite{Warren} was fitted to each $Q$ cut, and the obtained powder-averaged dispersion relation whose shape is sinusoidal is plotted in Fig.~\ref{Fig:LowE}(d).

\begin{figure}[t]
\centering
\includegraphics[width=8.54cm]{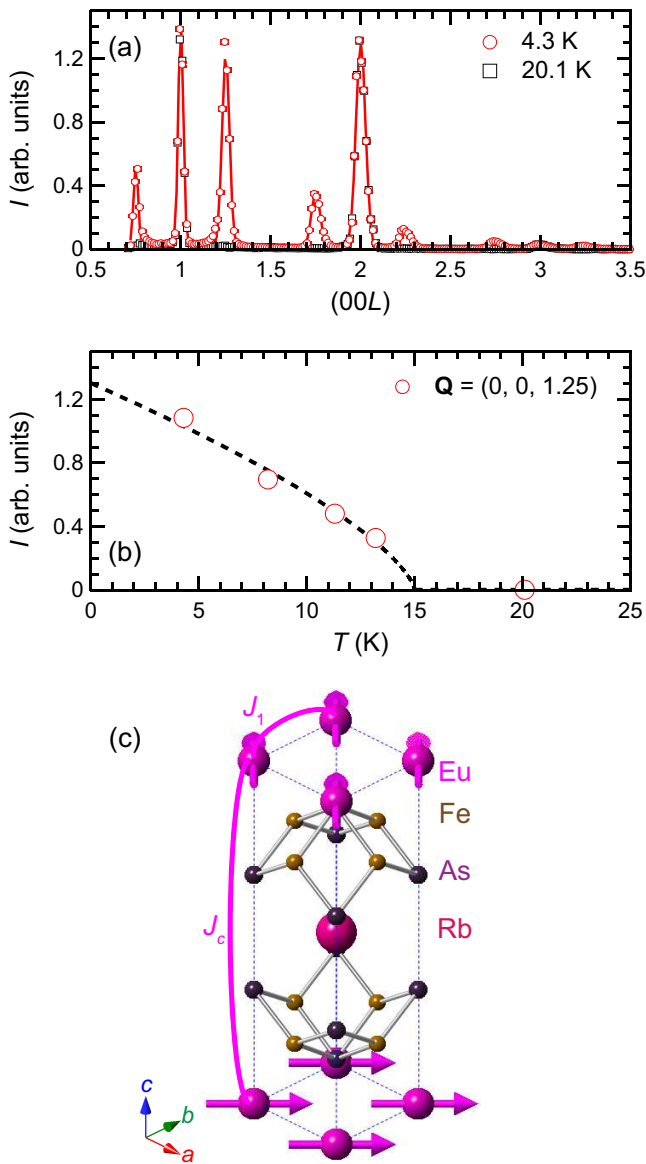}
\caption{
(a) Single-crystal neutron diffraction patterns along $(00L)$ in EuRbFe$_4$As$_4$ at $T=4.3$ and 20.1~K measured at SENJU.
Solid line is the fitting result using the magnetic propagation vector $\mathbf{k}=(0, 0, 0.25)$. 
(b) Temperature dependence of the neutron scattering intensity of the magnetic Bragg reflection at $\mathbf{Q}=(0,0,1.25)$ in EuRbFe$_4$As$_4$.
Dashed line is guide to the eye.
(c) Magnetic structure of the Eu sublattice in EuRbFe$_4$As$_4$ with the magnetic propagation vector $\mathbf{k}=(0, 0, 0.25)$.
Dashed lines represent the chemical unit cell.
The exchange couplings $J_1$ and $J_c$ are also described.
}
\label{Fig:Diffraction}
\end{figure}

To understand the spin-wave excitations from the Eu sublattice, the magnetic structure of Eu should be determined.
We now discuss the magnetic structure of the Eu sublattice in EuRbFe$_4$As$_4$.
Easy-plane magnetic anisotropy of the Eu$^{2+}$ moments in EuRbFe$_4$As$_4$ is reported by the single-crystal magnetic susceptibility~\cite{EuRbFe4As4_7} and heat capacity~\cite{EuRbFe4As4_9} measurements as in EuFe$_2$As$_2$~\cite{EuFe2As2_8,EuFe2As2_9}.
Furthermore, the recent M\"{o}ssbauer spectroscopy study on EuRbFe$_4$As$_4$ also reported that the Eu moments are ferromagnetically aligned within the $ab$ plane~\cite{EuRbFe4As4_4}.
Therefore, the remaining issue on the determination of the magnetic structure of the Eu sublattice in EuRbFe$_4$As$_4$ is how the two-dimensional in-plane ferromagnetic layers stack along the $c$ axis.
Magnetic Bragg reflections along the $(00L)$ direction give crucial information to address the issue.
As such, single-crystal neutron diffraction measurements along $(00L)$ below and above $T_\text{N}$ were performed.
As shown in Fig.~\ref{Fig:Diffraction}(a), at 4.3~K, several magnetic Bragg peaks were observed as satellite peaks beside the nuclear Bragg peaks with integer $L$, and the temperature dependence of the neutron scattering intensity of the magnetic Bragg peak at $\mathbf{Q}=(0,0,1.25)$ clearly represents the evolution of the long-rang magnetic order below $T_\text{N}=15$~K [Fig.~\ref{Fig:Diffraction}(b)].
Magnetic Bragg reflections appear at $L=0.75$, 1.25, 1.75, 2.25, 2.75, and 3.25, yielding the magnetic propagation vector $\mathbf{k}=(0,0,0.25)$.
As described by the solid line in Fig.~\ref{Fig:Diffraction}(a), the neutron diffraction pattern along $(00L)$ at 4.3~K is well fitted by considering $\mathbf{k}=(0,0,0.25)$; the amplitude and width of the Gaussian function for each peak as well as background are varied.
It should be noted that the neutron diffraction results along $(00L)$ can only determine the relative angle between the adjacent Eu layers but not the in-plane direction of the Eu moments.
For determination of the in-plane anisotropy in EuRbFe$_4$As$_4$, the magnetization curves along $\langle100\rangle$ and $\langle110\rangle$ are measured at 2~K as shown in Supplemental Material~\cite{Supplement}.
The $\langle110\rangle$ direction seems to be the easy axis, and the easy axis in EuRbFe$_4$As$_4$ is the same as that in EuFe$_2$As$_2$~\cite{EuFe2As2_6,EuFe2As2_7}.
The obtained magnetic structure is illustrated in Fig.~\ref{Fig:Diffraction}(c).
Three-dimensional antiferromagnetic order of the Eu sublattice can be seen, and the relative angle between Eu spins belonging to the nearest-neighbor layers is $90^{\circ}$.

Comparison between semi-classical LSW calculation and spin wave excitation provides us abundant information about the spin Hamiltonian~\cite{Lovesey}.
Here, we consider the following spin Hamiltonian for EuRbFe$_4$As$_4$: $\mathcal{H}=J_1\sum_{i,j}\mathbf{S}_i\cdot\mathbf{S}_j+J_c\sum_{i,j}\mathbf{S}_i\cdot\mathbf{S}_j$ where $\mathbf{S}$ is the spin operator of Eu$^{2+}$ ions ($S=7/2$).
$J_1$ and $J_c$ are the intra- and inter-plane nearest neighbor exchange coupling constants between Eu$^{2+}$ ions [Fig.~\ref{Fig:Diffraction}(c)].
The magnetic structure determined by our single-crystal neutron diffraction measurements illustrated in Fig.~\ref{Fig:Diffraction}(c) is considered.
Since $J_c$ is expected to be much smaller than $J_1$, we used the fixed value for $J_c$ (0.08~K) determined by the single crystal magnetization measurements~\cite{EuRbFe4As4_7}.
Fitting calculated powder-averaged dispersion relation to the experimental result as shown in Fig.~\ref{Fig:LowE}(d) yields the intra-plane ferromagnetic exchange constant $J_1/k_\text{B}=-1.31(1)$~K.
Similar intra-plane exchange coupling ($J_1/k_\text{B}\sim-1$~K) was reported in EuFe$_2$As$_2$ based on the local spin density approximation with the Coulomb repulsion (LSDA+$U$)~\cite{EuFe2As2_2}.
Calculated LSW intensity map in Fig.~\ref{Fig:LowE}(c) reproduces the experiment [Fig.~\ref{Fig:LowE}(a)] well; we mention that the LSW spectra based on $J_c=0.08$~K and 0 are almost the same as shown in Figs.~\ref{Fig:LowE}(c) and Supplemental Material~\cite{Supplement}.
Mean field approximation theory can estimate the Curie-Weiss temperature by the equation $\theta=-(z_1J_1+z_cJ_c)S(S+1)/3k_\text{B}$.
Here, the number of the nearest neighbor bonds $z_1$ and $z_c$ are 4 and 2 [Fig.~\ref{Fig:Diffraction}(c)].
By using $J_1$ obtained by our LSW analysis and $J_c$~\cite{EuRbFe4As4_7}, we estimate $\theta=26.7$~K in good agreement with the reported value (23.6~K) estimated by the susceptibility measurements~\cite{EuRbFe4As4_2}.
This result further corroborates the validity of our LSW analysis.
Important to be noted is that the further distance interaction along $c$ is required to account for the obtained magnetic structure [Fig.~\ref{Fig:Diffraction}(c)] which is not the conventional $A$-type antiferromagnetic order.
Since the further distance interaction along $c$ is expected to be very small, it is difficult to determine such interaction unless measuring a single crystal, which is left for future work.

Finally, we discuss the coupling between the long-range magnetic order of the Eu sublattice and the superconductivity of the FeAs layers in EuRbFe$_4$As$_4$.
$Q$ and $\hbar\omega$ dependences of the neutron spin resonance below and above $T_\text{N}$ are compared in Figs.~\ref{Fig:HighE}(b) and \ref{Fig:HighE}(c).
Onset of the long-range order of the Eu sublattice does not induce any noticeable change in both $Q$ and $\hbar\omega$ dependences of the neutron spin resonance mode within the accuracy of our INS measurements.
It should be noted that $T_\text{c}$ in EuRbFe$_4$As$_4$ is the highest among the 1144 systems although other 1144 compounds contain no local magnetic moment~\cite{CaKFe4As4_powder,EuRbFe4As4_1,EuRbFe4As4_2}.
EuRbFe$_4$As$_4$ do not exhibit the reentrant superconductivity below $T_\text{N}$~\cite{EuRbFe4As4_1,EuRbFe4As4_2} as mentioned above.
In Ni-doped EuRb(Fe$_{1-x}$Ni$_x$)$_4$As$_4$, the magnetic transition temperature of the Eu sublattice is independent of the Ni content $x$ although $T_\text{c}$ decreases rapidly~\cite{EuRbFe4As4_3}.
Moreover, substitution of nonmagnetic Ca decreases $T_\text{N}$ while $T_\text{c}$ is independent in the (Eu$_{1-x}$Ca$x$)RbFe$_4$As$_4$ system~\cite{EuRbFe4As4_12}.
Therefore, we conclude that the long-range magnetic order in the Eu sublattice plays a minor role in the superconductivity of the FeAs layers in EuRbFe$_4$As$_4$.
This may be reasonable because the Eu-$4f$ bands locate $2.7$~eV below the Fermi energy in EuRbFe$_4$As$_4$~\cite{EuRbFe4As4_10} as in doped EuFe$_2$As$_2$ systems~\cite{EuFeIr2As2_1,EuFeIr2As2_3,EuFe2As2_2,EuKFe2As2_2,EuFe2AsP2_7}.

\section{Conclusion}
In this work, we investigate both the static and dynamic magnetism in EuRbFe$_4$As$_4$.
On one hand, the spin resonance studied by INS measurements and PRA calculations in conjunction with the optical conductivity measurements~\cite{EuRbFe4As4_8} presents the strong evidence in favor of the sign-reversed $s_\pm$ superconductivity.
Our results show that EuRbFe$_4$As$_4$ belongs to the same paradigm carrying $s_\pm$ superconductivity as other iron pnictide systems although there is a static field from the long-range order of Eu$^{2+}$.
On the other hand, neutron diffraction measurements and LSW analysis determined the magnetic structure and the spin Hamiltonian of the Eu sublattice.
No signature of coupling between the long-range magnetic order of the Eu sublattice and the superconductivity of the FeAs layers in EuRbFe$_4$As$_4$ was observed in the $(Q, \hbar\omega)$ region probed in the current measurements.
To our best knowledge, this is the first INS study on Eu-containing iron-based superconductors for investigating the superconducting pairing symmetry and possible interplay between superconductivity and long-range magnetic order of Eu.

\section*{Acknowledgements}
We thank Ryoichi Kajimoto for helpful discussion.
Magnetic susceptibility measurements were performed at the CROSS user laboratories.
This research at SNS was sponsored by the Scientific User Facilities Division, Office of Basic Energy Sciences, U.S. Department of Energy.
Travel expenses for the ARCS experiment were provided by the General User Program for Neutron Scattering Experiments, Institute for Solid State Physics, The University of Tokyo (GPTAS:16913), at JRR-3, Japan Atomic Energy Agency (JAEA), Tokai, Japan.
The experiments at 4SEASONS and SENJU in J-PARC were conducted under the user program with the proposal numbers 2017A0005 and 2018B0008, respectively.
The calculations were performed by the supercomputing systems SGI ICE X at JAEA.
The present work was partially supported by JSPS KAKENHI Grant Numbers JP15K17712, JP17K14349, and JP18K03552, and the Cooperative Research Program of ``Network Joint Research Center for Materials and Devices'' (20181072).
ADC was partiality supported by the U.S. DOE, Office of Science, Basic Energy Sciences, Materials Sciences and Engineering Division.
DK and ME were supported by the Austrian Science Fund (FWF) I2814-N36.

\end{document}